# DAFNet: Dual-path Adaptive Fusion Network for High-Fidelity Cross-Modality Brain MR Image Synthesis

Jialin Liu[1], Xiaoliang Zhang[1,2]*

[1]Department of Biomedical Engineering, [2]Department of Electrical Engineering, State University of New York at Buffalo, Buffalo, New York 14260, U.S.A.

*Corresponding author:

Xiaoliang Zhang, Ph.D.
Bonner Hall 215E
Department of Biomedical Engineering
State University of New York at Buffalo
Buffalo, NY 14260
U.S.A.

Email: xzhang89@buffalo.edu

**Abstract:** Cross-modality MRI synthesis using deep learning can streamline clinical workflows, compensate for missing image contrasts, and reduce examination time; however, existing methods often require substantial computational resources, limiting their practical accessibility. To address this limitation, we propose and investigate a lightweight yet effective framework, termed the Dual-path Adaptive Fusion Network (DAFNet), for T1-to-T2 images synthesis. DAFNet is based on cGAN (conditional generative adversarial networks) structure, and the generator employs a dual-path encoder that adaptively combines depth-wise separable convolutions and dilated convolutions to capture both fine-grained details and broader contextual features. A novel Dual-path Adaptive Fusion Module (DAFM) is introduced to dynamically combine these feature streams through channel-wise complementary weighting, enabling efficient and adaptive feature integration. This fusion mechanism is further applied to the final skip connection to enhance reconstruction fidelity, while the decoder adopts a standard transposed convolution architecture. Coupled with a conventional conditional GAN discriminator, the proposed model maintains low computational complexity while improving perceptual image fidelity. Experimental results demonstrate that DAFNet achieves superior performance compared with baseline U-Net and conventional cGAN models, reaching a peak signal-to-noise ratio (PSNR) of 26.43 dB while significantly reducing model size. This result indicates that DAFNet provides an effective balance between synthesis fidelity and computational efficiency, making it a promising solution for deployment on standard computing hardware and for broader clinical and research applications in MRI contrast synthesis.

# 1. Introduction

Magnetic resonance imaging (MRI) [1, 2] provides a wide range of tissue contrasts, each offering distinct clinical and research value. Among the most commonly used contrasts, T1-weighted (T1w) images provide detailed anatomical information and clear gray–white matter delineation, whereas T2-weighted (T2w) images are particularly sensitive to fluid content, edema, and various pathological changes. These complementary properties make multi-contrast MRI valuable for comprehensive tissue characterization and clinical diagnosis [3, 4]. However, acquiring multiple image contrasts requires separate pulse sequences, leading to prolonged examination time, increase cost, and greater susceptibility to motion artifacts [5]. Considerable efforts have therefore been devoted to accelerating MRI acquisition [6-17]. Parallel imaging reduces acquisition time by undersampling k-space and using the spatial sensitivity profiles of multichannel receiver arrays to reconstruct the missing information [18-21], whereas compressed sensing exploits image sparsity and incoherent undersampling to recover images from substantially reduced k-space data [22-25]. Although these techniques can significantly accelerate MRI, higher acceleration factors are generally accompanied by penalties in signal-to-noise ratio (SNR), reconstruction quality, or both [26-28]. The intrinsically higher SNR available at high and ultrahigh magnetic field strengths can provide additional SNR [29-37] reserve for accelerated imaging; however, higher field strengths also introduce technical challenges [38-56] and are not universally available. Thus, despite major advances in acquisition acceleration, obtaining multiple MR contrasts within a short examination remains an important practical challenge.

Cross-modality MRI synthesis provides a fundamentally different and complementary approach to this problem by computationally generating a desired image contrast from one or more already acquired contrasts [57]. Rather than solely accelerating the acquisition of each individual sequence, synthesis of a clinically useful contrast from an existing acquisition has the potential to eliminate the need for an additional sequence in selected applications, thereby further reducing examination time and providing missing contrasts when an acquisition is unavailable or unusable. In recent years, deep learning methods have demonstrated considerable potential for cross-modality MRI synthesis. Convolutional neural networks (CNNs), particularly U-Net–based architectures[58], have been widely adopted because their encoder–decoder structures and skip connections facilitate the preservation of spatial information. However, conventional CNN-based approaches may produce overly smooth images and may have difficulty recovering realistic textures and fine anatomical details.

To improve perceptual image quality, generative adversarial networks (GANs)[59] have been introduced for MRI synthesis. Conditional GAN (cGAN) frameworks [60, 61], such as pix2pix, incorporate adversarial training to encourage the generation of visually realistic images. Other approaches, including CycleGAN[62, 63], further enable cross-modality translation without requiring strictly paired training data. Despite these advances, GAN-based approaches may suffer from training instability and can involve relatively complex network architectures, resulting in substantial computational and memory requirements. Such requirements may limit their practical accessibility, particularly in clinical or research environments where high-performance computing resources are not readily available.

To address these challenges, we propose a lightweight and efficient deep learning framework, termed the Dual-path Adaptive Fusion Network (DAFNet), for T1-to-T2 brain MRI synthesis. The core of DAFNet is a dual-path encoder designed to extract complementary image features using depthwise separable dilated convolutions and depthwise separable convolutions . The dilated-convolution pathway provides an enlarged receptive field for capturing broader contextual information, whereas the complementary pathway emphasizes local and fine-grained structural features. To effectively integrate information from these two pathways, we introduce a novel Dual-path Adaptive Fusion Module (DAFM), which performs channel-wise adaptive fusion through a complementary weighting mechanism. This design allows the network to dynamically balance features from the two pathways while reducing model complexity compared with conventional concatenation-based feature fusion.

The adaptive fusion mechanism is incorporated throughout the encoder and is also applied to the final skip connection to enhance feature integration during image reconstruction. The DAFNet generator is paired with a conditional PatchGAN discriminator to further improve perceptual quality and the fidelity of the synthesized images through adversarial learning. By combining dual-path feature extraction, adaptive feature fusion, and a lightweight network architecture, DAFNet is designed to achieve a favorable balance between synthesis fidelity and computational efficiency [64, 65].

The proposed method was evaluated using a publicly available dataset of paired T1w and T2w brain MR images and compared with baseline U-Net and conventional cGAN models. DAFNet demonstrated improved quantitative and qualitative synthesis performance while substantially reducing the number of generator parameters. These characteristics suggest that DAFNet may provide an efficient approach to cross-modality MRI synthesis, with particular potential for applications in resource-constrained clinical and research environments where access to high-performance computing hardware is limited.

# 2. Methods

## 2.1 Dataset

The dataset used in this study was obtained from a publicly available resource released by the Adolescent Health and Development in Context (AHDC) study (OpenNeuro accession: ds005901)[66]. This longitudinal dataset contains structural MRI scans from 119 subjects, including paired T1-weighted (T1w) and T2-weighted (T2w) images.

All images are provided in NIfTI format. For each subject, T1w and T2w images are stored within the same directory under an "anat" subfolder, ensuring consistent pairing between modalities. Only structural T1w and T2w images were used in this study. Detailed acquisition parameters, participant demographics, and ethical considerations are available through the original dataset repository.

## 2.2 Deep Learning Model

The proposed Dual-path Adaptive Fusion Network (DAFNet) consists of a U-Net–based generator with a dual-path encoder and an adaptive feature fusion mechanism. The generator is paired with a conditional PatchGAN discriminator to enhance perceptual realism through adversarial training.

### 2.2.1 Dual-path Adaptive Fusion Module (DAFM)

The Dual-path Adaptive Fusion Module (DAFM) is designed to efficiently integrate features extracted from two parallel convolutional paths. Given two feature maps, the module first concatenates them along the channel dimension, followed by a convolutional layer to reduce the number of channels to one-eighth of the original size. The model structure is shown in **Figure 1**.

To capture global contextual information efficiently, spatial compression is applied through pooling operations to reduce feature dimensions. The resulting representation is processed through convolutional and fully connected layers to generate a compact feature vector. A sigmoid activation function is then applied to produce channel-wise weights in the range of [0, 1].

Feature fusion is performed using a complementary weighting mechanism: one feature map is scaled by the learned weights, while the other is scaled by the complementary weights (1 − w). The final output is obtained through element-wise weighted summation.

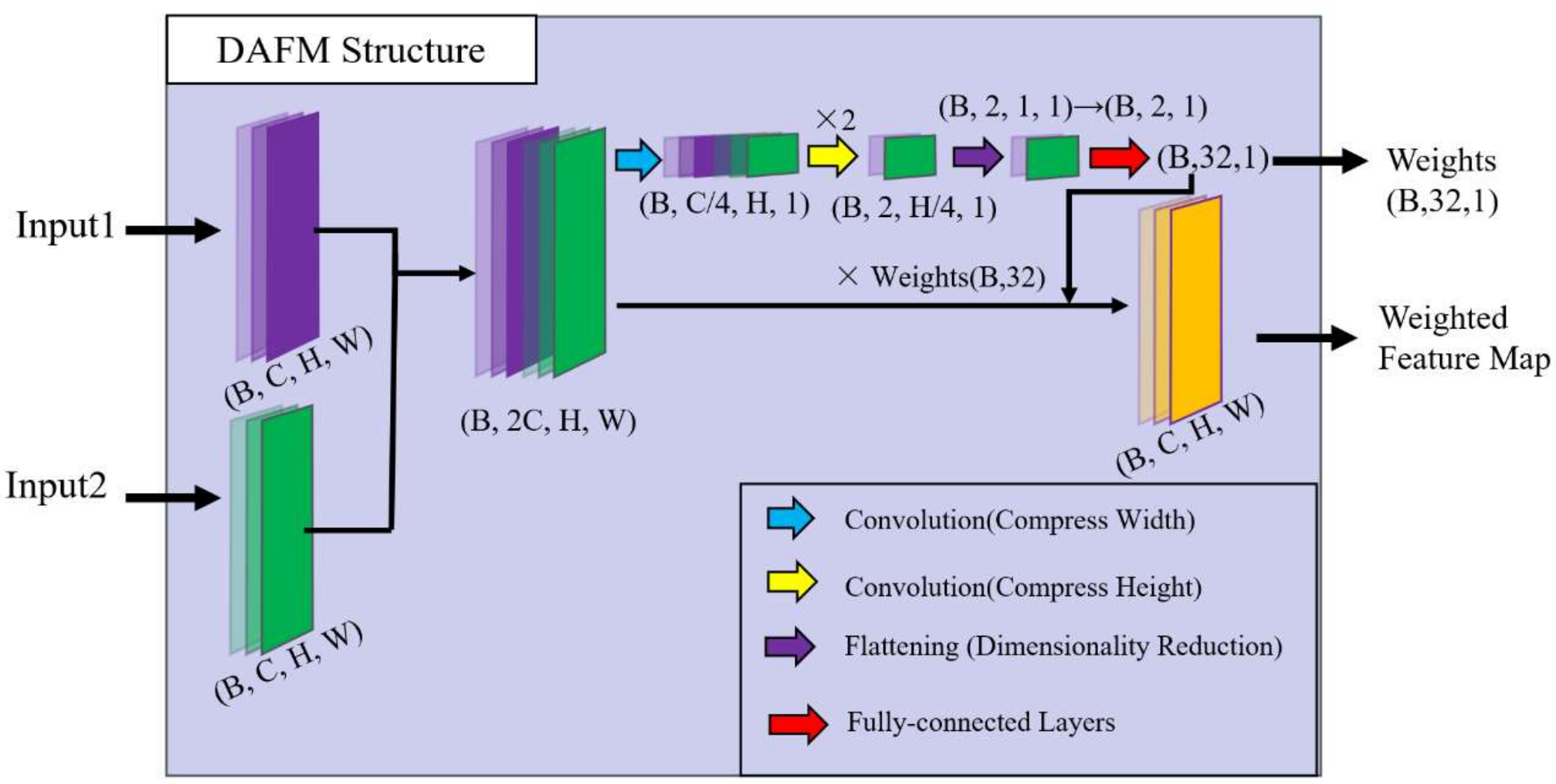

**Figure 1** Structure of the Dual-path Adaptive Fusion Module (DAFM). The module adaptively fuses features from two parallel paths using channel-wise complementary weighting based on global feature representation.

This design enables adaptive, channel-wise feature selection while avoiding the parameter overhead of traditional concatenation-based fusion. By leveraging global feature information, the module dynamically balances the contribution of each path, allowing the network to capture both large-scale contextual structures and fine-grained details, while maintaining a lightweight architecture.

### 2.2.2 U-Net Generator Architecture

The generator in DAFNet is based on a U-Net architecture with a dual-path encoder and a standard decoder. Both the encoder and decoder consist of four layers. As shown in **Figure 2**.

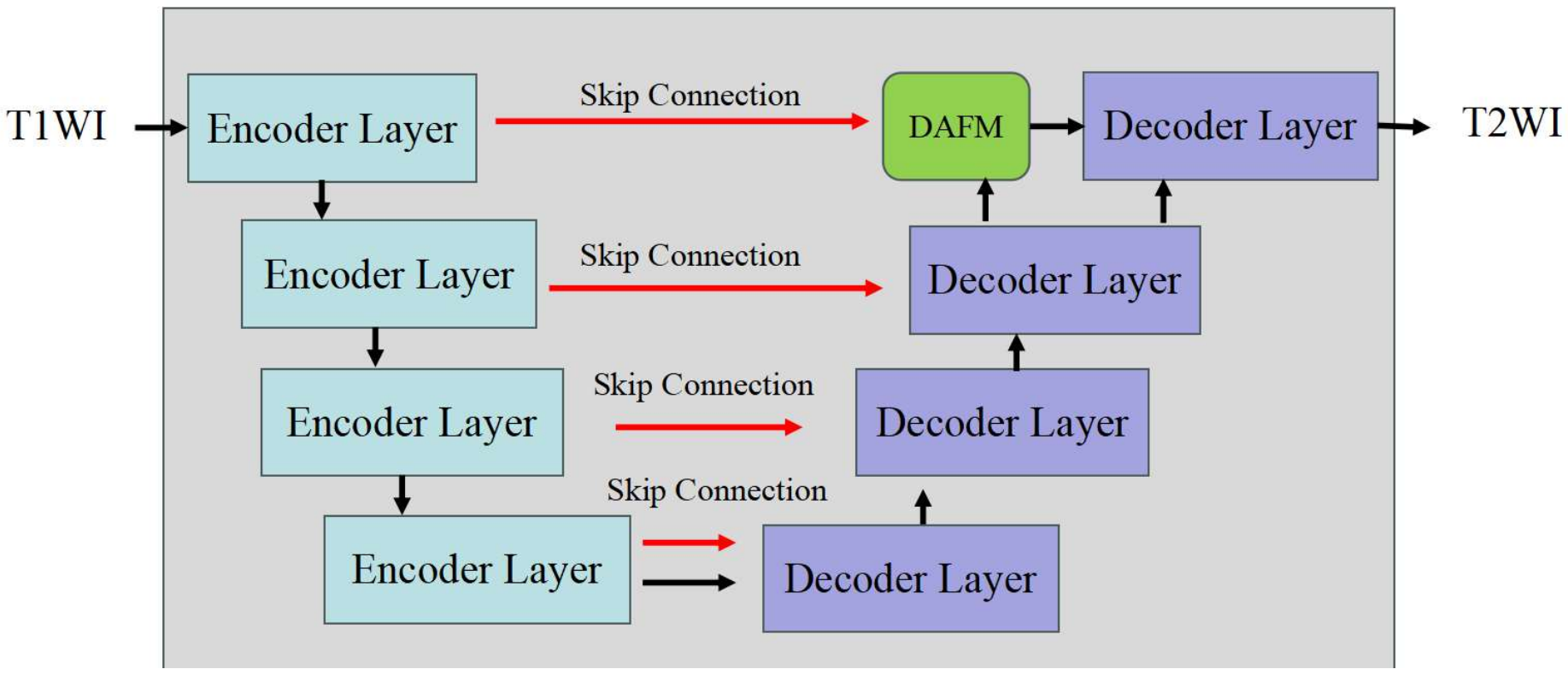

**Figure 2** Overall architecture of the proposed DAFNet. The framework consists of a U-Net–based generator with a dual-path encoder and adaptive fusion modules, paired with a conditional PatchGAN discriminator.

In the encoder, each layer contains two parallel convolutional paths. As shown in **Figure 3**. The first path employs depthwise separable dilated convolutions (kernel size 3×3) to capture features with a large receptive field, while the second path uses a combination of standard convolution and depthwise separable convolution to extract fine structural details. The outputs of these two paths are fused using the DAFM.

The fused feature maps are passed through pooling and activation layers and propagated to subsequent layers. Skip connections are preserved to retain high-resolution spatial information. The decoder consists of four layers. As shown in **Figure 4**. Each of the first three layers performs transposed convolution for upsampling, followed by concatenation with the corresponding encoder features via skip connections. The combined features are processed using convolutional layers with batch normalization and ReLU activation.

In the final decoder layer, the DAFM is applied to fuse the upsampled feature map with the corresponding skip connection from the first encoder layer. A final 1×1 convolution produces the synthesized T2-weighted image.

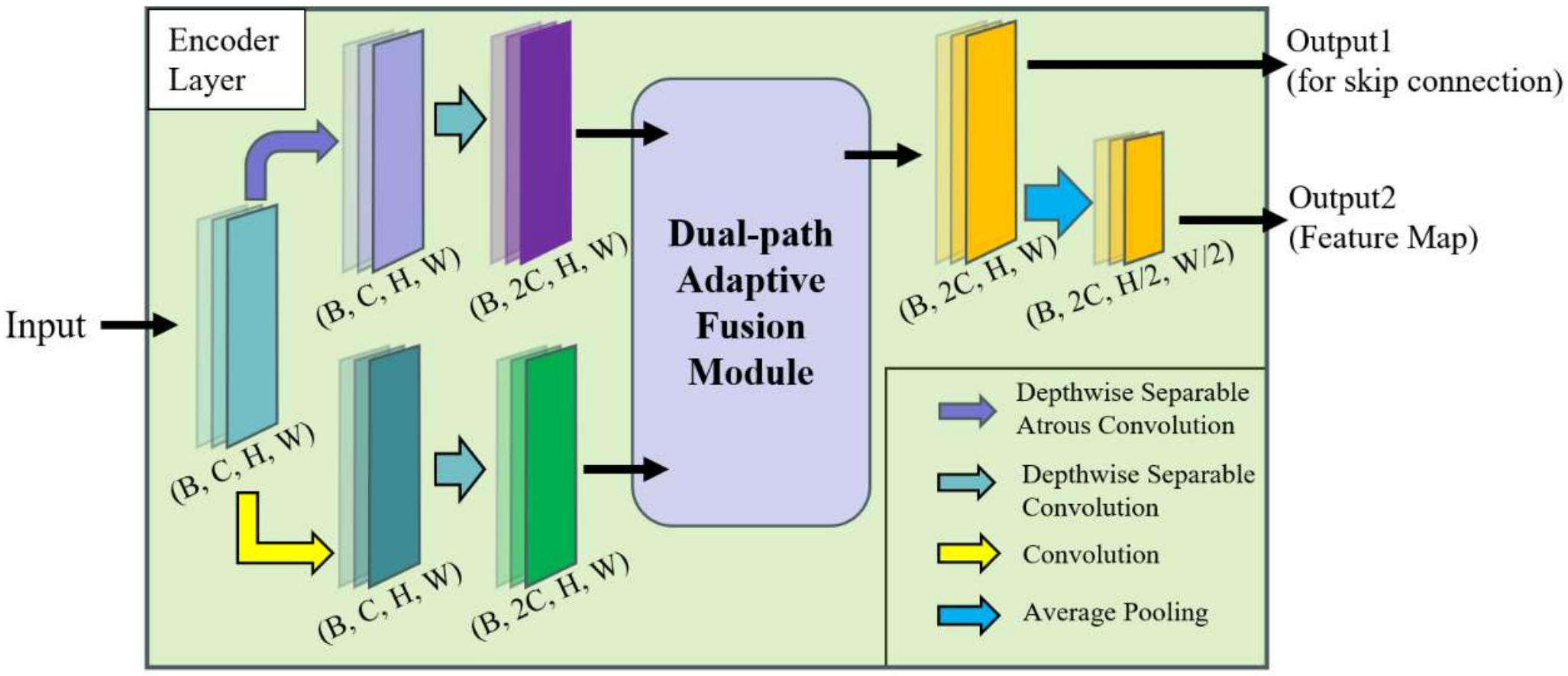

**Figure 3** Structure of an encoder layer in DAFNet. Each layer includes two parallel convolutional paths for multi-scale feature extraction, followed by adaptive fusion using the DAFM.

### 2.2.3 Conditional PatchGAN Discriminator Architecture

The proposed model employs a conventional conditional PatchGAN discriminator. As shown in **Figure 5**. It processes the input and target image through separate initial convolution branches. A notable architectural addition is the integration of the same DAFM within each of its middle layers. In each layer, features from the two input branches are adaptively fused using this module, generating channel-wise weights to guide their combination, before being passed to the next layer. This design enables dynamic, data-driven fusion of conditional and generated information at multiple feature scales. The final layer outputs a patch-based discrimination map to guide the generator by calculating the adversarial loss based on this discrimination map.

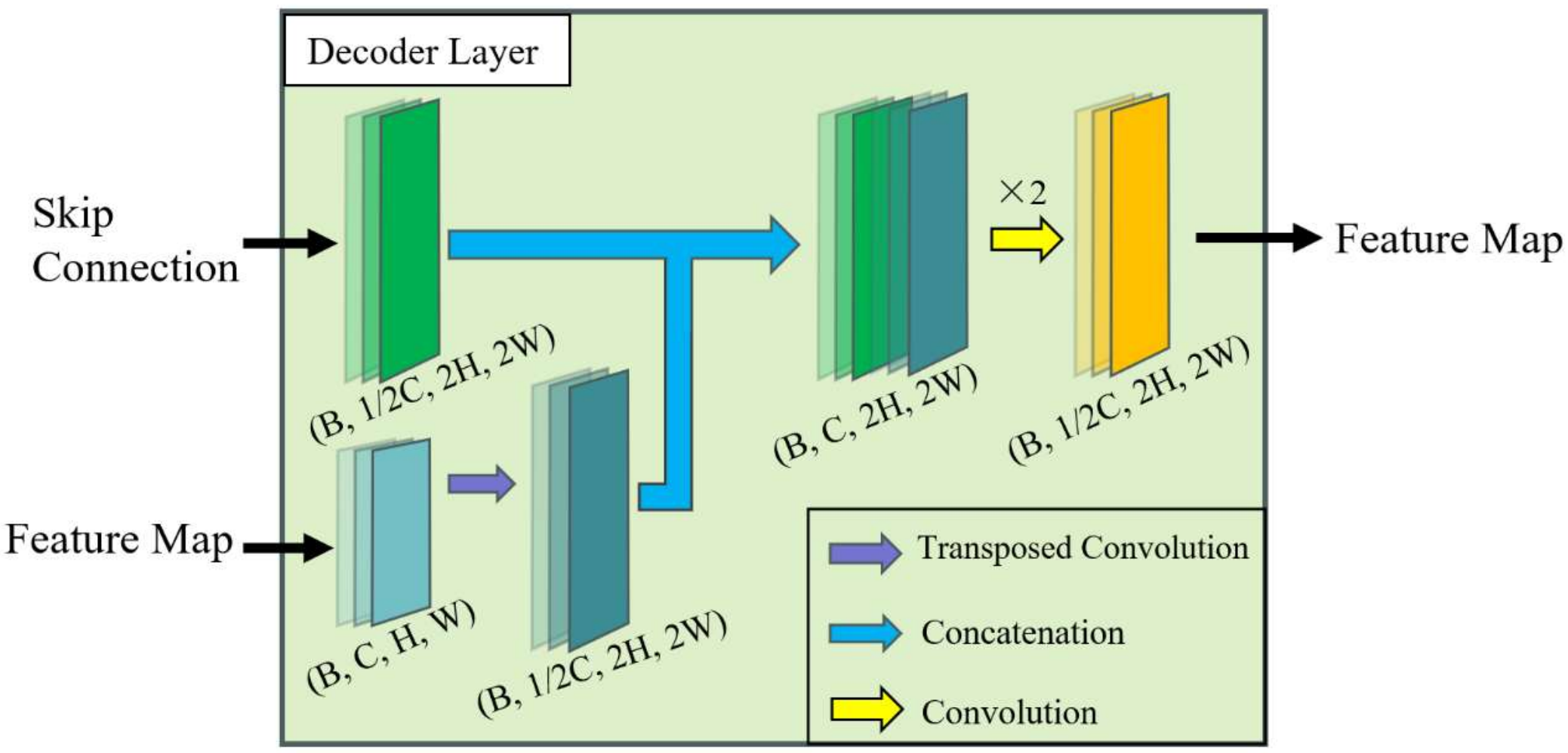

**Figure 4** Structure of a decoder layer in DAFNet. The decoder performs upsampling via transposed convolution and integrates skip connections from the encoder, with adaptive fusion applied in the final layer.

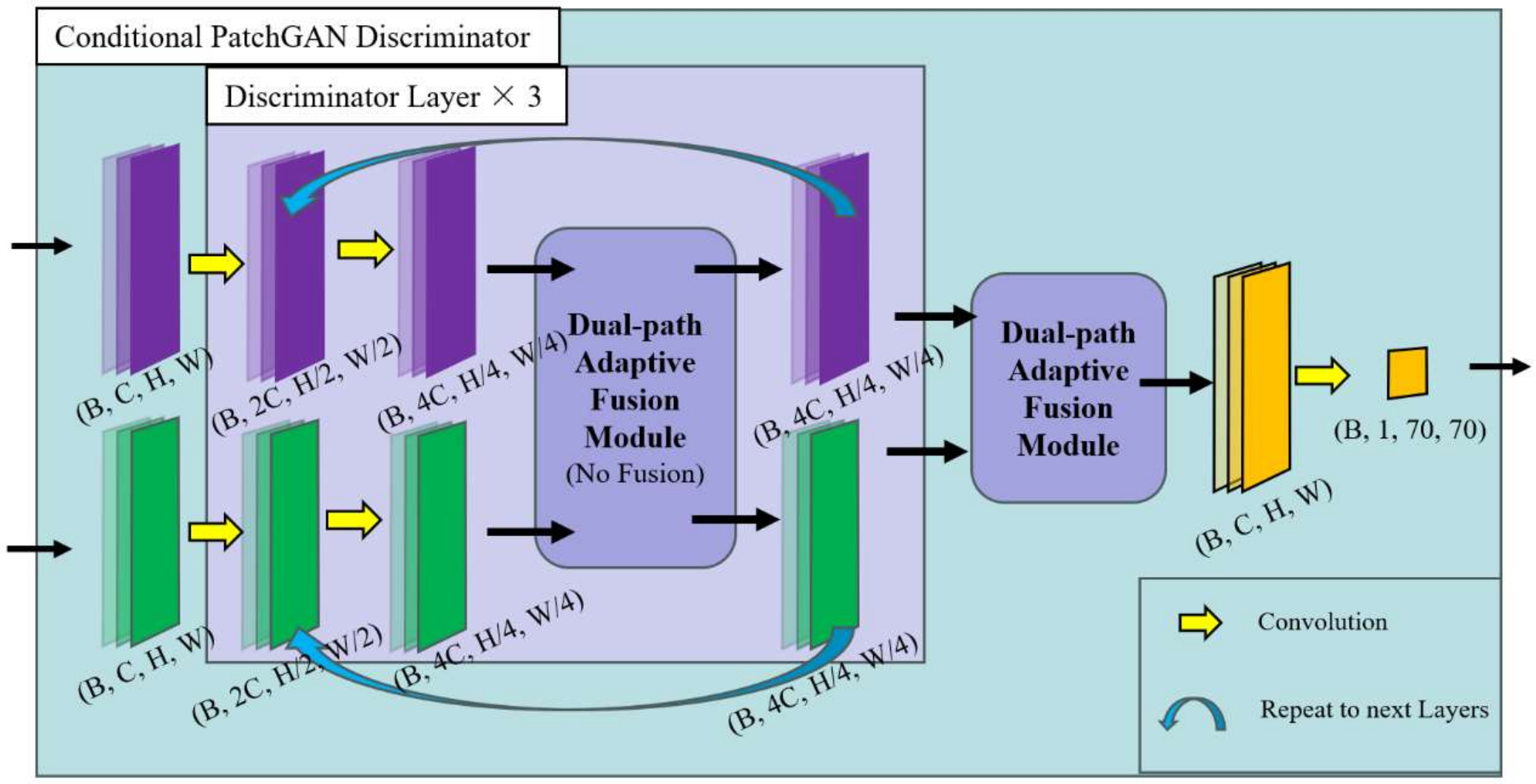

**Figure 5** Architecture of the conditional PatchGAN discriminator. The discriminator processes paired inputs and patch-wise classification, with adaptive fusion applied at intermediate layers to integrate conditional and generated features.

2.2.4 Loss function

A two-phase training strategy was employed to improve stability and reconstruction performance.

During the pre-training phase (first 8 epochs), only the generator was optimized using a weighted combination of reconstruction losses, including mean squared error (MSE, weight = 20), structural similarity index (SSIM[67], weight = 0.3), L1 loss (weight = 15), and a contrast enhancement loss (weight = 50). Pixel-value weighting was applied to emphasize anatomically relevant intensity ranges, which stabilizes optimization and improves feature learning in critical regions. No adversarial loss was used in this phase.

Starting from epoch 9, adversarial training was introduced. The generator loss was defined as the weighted sum of the reconstruction loss and an adversarial loss with an initial weight of 0.1. The adversarial component was implemented using a least-squares GAN (LSGAN) formulation.

The discriminator was trained using both real image pairs (T1w + real T2w) and generated pairs (T1w + synthesized T2w), with label smoothing applied (target values of 0.9 for real samples and 0.1 for generated samples).

An adaptive training strategy was used to improve convergence. If either the generator or discriminator validation loss did not improve for 5 consecutive epochs, a corresponding early-stopping counter was incremented. When the discriminator counter reached 5, its learning rate and adversarial loss weight were reduced by half, and the counter was reset. When the generator counter reached 5, training was terminated and the best-performing model was saved. Model checkpoints were saved every 5 epochs.

The SSIM loss was implemented as (1 − SSIM) to ensure consistency with loss minimization objectives.

# 3. Experiments

## 3.1 Preprocessing

All preprocessing steps were applied consistently to ensure spatial alignment and data uniformity. T1-weighted and T2-weighted images were first separated into modality-specific groups while preserving subject-level pairing.

All volumes were cropped to a uniform size of $256 \times 256 \times 64$. Cropping was performed by removing low-intensity background regions from the image boundaries, and identical cropping coordinates were applied to both T1w and T2w images to maintain spatial correspondence.

The dataset was split at the subject level, with 100 subjects used for training and 19 subjects used for testing.

Data augmentation was applied to the training set only. Each volume underwent a random rotation of $\pm 10°$ along the longitudinal axis to improve generalization.

Intensity normalization was performed by scaling voxel values from the original range [0, 4095] to [0, 1], which stabilizes optimization and prevents gradient instability during training.

To reduce computational cost, 3D volumes were decomposed into 2D slices. Axial slices were extracted from each volume, resulting in multiple training samples per subject while significantly reducing memory requirements.

## 3.2 Training Strategy

.A two-phase training strategy was adopted to improve stability and reconstruction performance.

During the pre-training phase (first 8 epochs), only the generator was optimized using a weighted combination of pixel-level reconstruction losses, including mean squared error (MSE, weight = 20), structural similarity index (SSIM, weight = 0.3), L1 loss (weight = 15), and a contrast enhancement loss (weight = 50). Pixel-value weighting was applied to emphasize anatomically relevant intensity ranges, which stabilizes optimization and improves feature learning in critical regions. No adversarial loss was used in this phase.

Starting from epoch 9, adversarial training was introduced. The generator loss was defined as the weighted sum of the reconstruction loss and an adversarial loss with an initial weight of 0.1. The adversarial component was implemented using a least-squares GAN (LSGAN) formulation.

The discriminator was implemented as a conditional PatchGAN and trained using both real image pairs (T1w + real T2w) and generated pairs (T1w + synthesized T2w). Label smoothing was applied, with target values of 0.9 for real samples and 0.1 for generated samples, to improve training stability.

An adaptive training strategy was employed to further enhance convergence. If either the generator or discriminator validation loss did not improve for 5 consecutive epochs, a corresponding early-stopping counter was incremented. When the discriminator counter reached 5, its learning rate and the adversarial loss weight were reduced by half, and the counter was reset. When the generator counter reached 5, training was terminated and the best-performing model was saved. Model checkpoints were also saved every 5 epochs.

# 4. Results

## 4.1 Qualitative Evaluation

Visual comparisons of synthesized T2-weighted images are presented in **Figure 6** and **Figure 7**. Compared with baseline U-Net and conventional cGAN models, the proposed DAFNet demonstrates improved structural fidelity and visual realism.

Specifically, images generated by the U-Net model appear noticeably blurred, with reduced contrast between different brain tissues and poorly defined anatomical boundaries. The cGAN model improves overall sharpness; however, it introduces excessive contrast in certain regions, leading to unrealistic signal amplification, particularly in white matter structures.

In contrast, DAFNet produces images with clearer anatomical boundaries, more consistent tissue contrast, and fewer artifacts. The synthesized images more closely resemble the ground-truth T2-weighted images in both structural detail and intensity distribution.

As illustrated in **Figure 7**, DAFNet achieves superior reconstruction in challenging regions such as the orbital area, where it preserves fine structural details that are either blurred or distorted in the baseline methods.

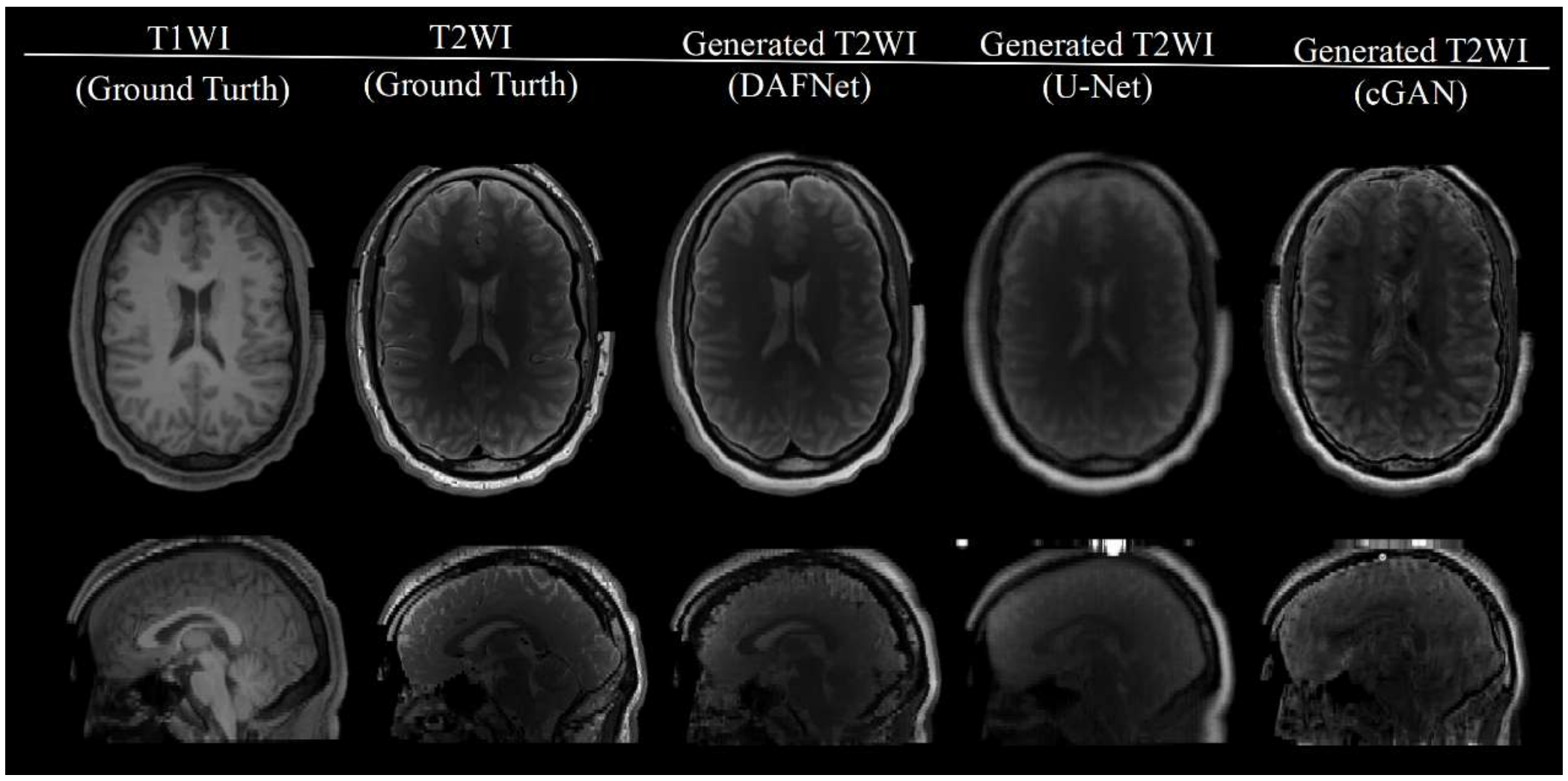

**Figure 6** Comparison of synthesized T2-weighted MRI images from different models (case: s116). Representative results from U-Net, cGAN, and DAFNet are shown alongside the ground-truth T2-weighted image. DAFNet demonstrates improved structural fidelity, sharper tissue boundaries, and more realistic contrast compared with baseline methods.

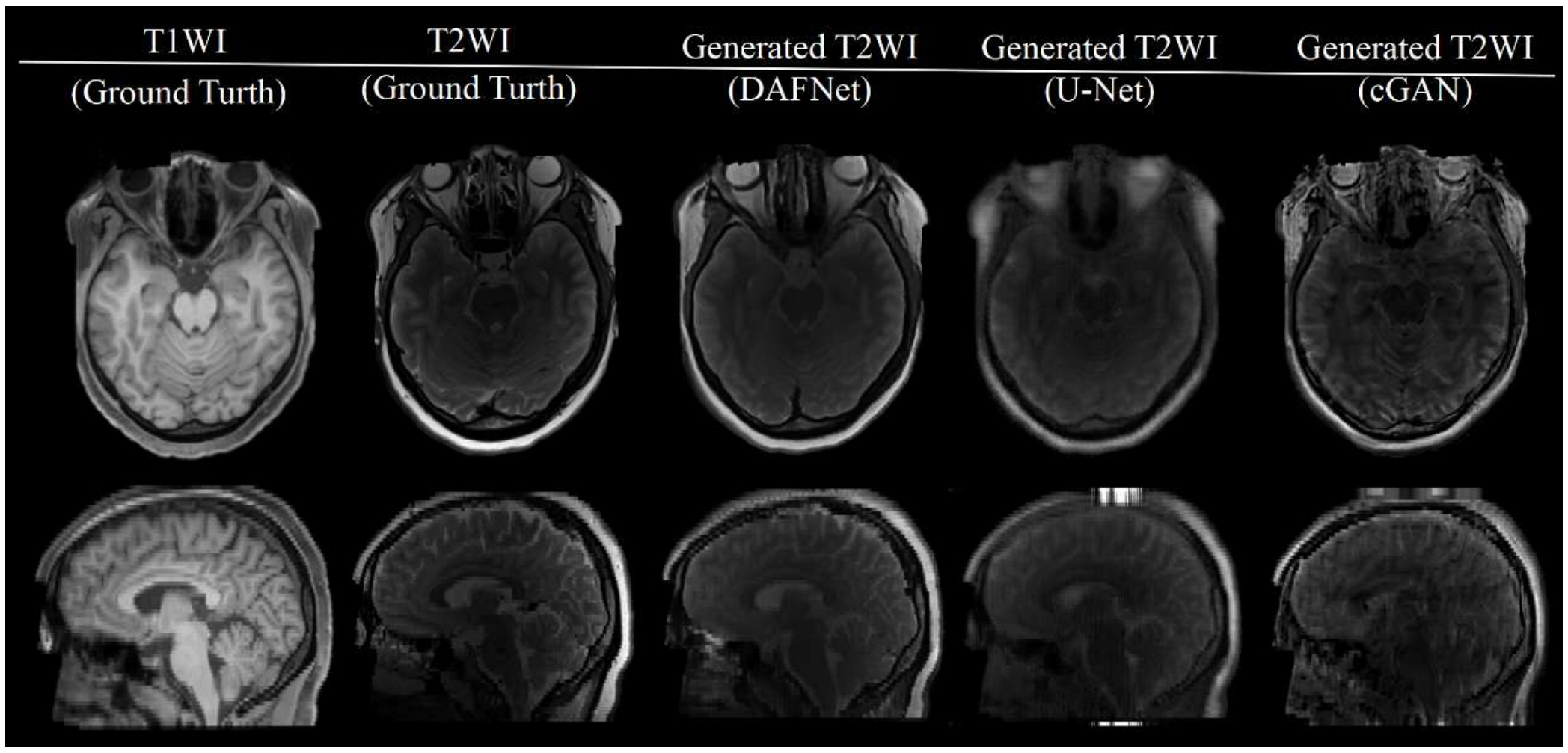

**Figure 7** Comparison of synthesized T2-weighted MRI images from different models (case: s120). In the orbital region, DAFNet preserves fine anatomical structures and reduces artifacts, whereas U-Net and cGAN exhibit blurring and structural distortion.

**Table 1** Quantitative evaluation of MRI synthesis performance. Comparison of peak signal-to-noise ratio (PSNR), structural similarity index (SSIM), and mean squared error (MSE) for U-Net, cGAN, and the proposed DAFNet. Values are reported as mean ± standard deviation. “G” and “D” denote the number of parameters in the generator and discriminator, respectively.

| Methods | PSNR | SSIM | MSE | # parameters |
|---|---|---|---|---|
| U-Net | 24.24 ± 0.41 | 0.7810 ± 0.231 | 0.0031 ± 0.0012 | 54.41M |
| cGANs | 25.33 ± 0.96 | 0.7918 ± 0.132 | 0.0027 ± 0.0061 | 54.41M(G) + 2.77M(D) |
| DAFNet | 26.43 ± 0.77 | 0.852 ± 0.065 | 0.0021 ± 0.0008 | 11.59M(G) + 7.28M(D) |

## 4.2 Quantitative Evaluation

Quantitative results are summarized in

**Table 1**. The proposed DAFNet achieves the best overall performance across all evaluation metrics.

Compared with the baseline U-Net model, DAFNet improves the peak signal-to-noise ratio (PSNR) from 24.24 ± 0.41 dB to 26.43 ± 0.77 dB, increases the structural similarity index (SSIM) from 0.7810 ± 0.231 to 0.852 ± 0.065, and reduces the mean squared error (MSE) from 0.0031 ± 0.0012 to 0.0021 ± 0.0008.

Compared with the cGAN model, DAFNet also demonstrates consistent improvements, increasing PSNR from 25.33 ± 0.96 dB to 26.43 ± 0.77 dB, improving SSIM from 0.7918 ± 0.132 to 0.852 ± 0.065, and reducing MSE from 0.0027 ± 0.0061 to 0.0021 ± 0.0008.

In addition to improved reconstruction accuracy, DAFNet significantly reduces model complexity. The number of generator parameters is reduced from 54.41M (U-Net) to 11.59M, representing a reduction of approximately 78.7%. Although the discriminator in DAFNet contains 7.28M parameters, the overall architecture remains computationally efficient compared with conventional GAN-based models.

These results demonstrate that the proposed dual-path adaptive fusion mechanism effectively improves synthesis performance while maintaining a lightweight model design.

### 4.3 Evaluation Metrics

The peak signal-to-noise ratio (PSNR) measures the similarity between synthesized and ground-truth images based on pixel intensity differences, with higher values indicating better reconstruction quality.

The mean squared error (MSE) quantifies the average squared difference between corresponding pixel values, where lower values indicate better performance.

The structural similarity index (SSIM) evaluates perceptual similarity by considering luminance, contrast, and structural information, with values closer to 1 indicating higher image quality.

# 5. Conclusion and Discussion

## 5.1 Conclusion

In this study, we developed DAFNet, a lightweight deep learning framework for cross-modality T1-to-T2 brain MRI synthesis. By incorporating the proposed Dual-path Adaptive Fusion Module (DAFM) into the dual-path encoder and the final skip connection of a U-Net–based generator, DAFNet enables efficient and adaptive integration of complementary image features. The dual-path architecture captures both fine-grained structural information and broader contextual features, while the DAFM dynamically balances their contributions through channel-wise complementary weighting.

Quantitative evaluation demonstrated that DAFNet consistently outperformed the baseline U-Net and conventional cGAN models across all evaluated image-quality metrics, including peak signal-to-noise ratio (PSNR), structural similarity index (SSIM), and mean squared error (MSE). DAFNet achieved a PSNR of 26.43 ± 0.77 dB, an SSIM of 0.852 ± 0.065, and an MSE of 0.0021 ± 0.0008. Qualitative evaluation further showed improved preservation of anatomical boundaries and tissue contrast, with fewer visible artifacts compared with the baseline methods.

Importantly, these improvements were achieved with a substantially smaller generator. The number of generator parameters was reduced by approximately 78.7%, from 54.41 million in the baseline U-Net and cGAN generators to 11.59 million in DAFNet. Thus, the proposed architecture improves synthesis fidelity while substantially reducing model size. This combination of reconstruction performance and model efficiency makes DAFNet a promising approach for cross-modality MRI synthesis, particularly for clinical and research environments in which access to high-performance computing resources may be limited.

## 5.2 Discussion

The results of this study demonstrate the potential of adaptive feature fusion for improving the efficiency and fidelity of cross-modality MRI synthesis. The performance of DAFNet suggests that increasing network size is not necessarily required to improve synthesis quality. Instead, efficiently extracting and integrating complementary features may provide a more effective strategy. In DAFNet, the two encoder pathways are designed to capture information at different spatial scales, while the DAFM adaptively determines their relative contributions. This design may help explain why DAFNet achieved improved quantitative and qualitative performance despite using substantially fewer generator parameters than the baseline models.

The improved synthesis performance may also be attributed to the complementary roles of the reconstruction and adversarial objectives. The reconstruction losses encourage voxel-wise and structural consistency with the reference T2w images, whereas the conditional PatchGAN discriminator promotes more realistic local image characteristics. The two-stage training strategy, in which the generator is first pretrained using reconstruction losses before adversarial training is introduced, was designed to improve training stability and reduce the likelihood that adversarial learning would dominate before the generator had established a reasonable T1w-to-T2w mapping.

Several limitations of the present study should nevertheless be considered. First, only minimal preprocessing was applied to the dataset, and no explicit image registration was performed between the paired T1w and T2w images. Consequently, residual spatial misalignment may remain between the two modalities. Because the network is trained using paired images and reconstruction losses that depend on spatial correspondence, such misregistration may interfere with learning of the underlying cross-modality relationship and may also influence quantitative metrics based on direct comparison between synthesized and reference images.

Second, intersubject anatomical variability and residual differences in spatial correspondence between T1w and T2w images may contribute to variations in synthesis quality. These factors could partially account for localized artifacts, reduced sharpness, or differences in intensity distribution observed in some synthesized images. Future studies incorporating image registration and more systematic background removal may improve spatial consistency between paired images and allow the effects of the proposed network architecture to be evaluated more accurately.

Third, although DAFNet substantially reduces the number of generator parameters, parameter count alone does not fully characterize computational efficiency. Other measures, including floating-point operations, memory consumption, and inference time on different hardware platforms, were not evaluated in the present study. These measurements will be important in future work to more comprehensively assess the computational advantages of DAFNet and its feasibility for deployment on standard computing hardware.

The present evaluation was also conducted using a single publicly available brain MRI dataset. Further validation using larger and more diverse datasets will therefore be necessary to determine the generalizability of the proposed approach across different populations, scanners, field strengths, acquisition protocols, and imaging centers. More extensive data augmentation and repeated data-splitting strategies, such as Monte Carlo cross-validation [68, 69], could provide a more robust assessment of model performance and reduce potential dependence on a particular training–test split.

Future work will focus on further optimization of the network and training strategy, including adaptive adjustment of the adversarial loss weight and discriminator complexity. Extension from the current 2D implementation to a 3D framework may also allow the network to exploit through-plane anatomical information and improve spatial consistency across adjacent slices, although this would increase computational and memory requirements. Finally, evaluation on multi-center datasets and direct assessment of inference time, memory requirements, and performance on standard computing hardware will be important for establishing the practical utility of the proposed lightweight architecture.

Overall, the present results demonstrate that DAFNet can achieve improved T1-to-T2 MRI synthesis with a substantially reduced generator size. The proposed dual-path architecture and adaptive fusion mechanism provide an efficient means of integrating complementary local and contextual features, offering a promising foundation for further development of computationally efficient cross-modality MRI synthesis.

**Authorship Contribution Statement**

Jialin Liu: Methodology, Software, Formal Analysis, Investigation, Resources, Data Curation, Writing – Original Draft, Visualization. Xiaoliang Zhang: Conceptualization, Methodology, Formal Analysis, Investigation, Resources, Writing – Editing, Validation, Supervision, Project Administration.